\ifdefined\WORDCOUNT
\documentclass[superscriptaddress,twocolumn,preprintnumbers,amsmath,amssymb,prl,nofootinbib,groupedaddress]{revtex4-1}
\else
\documentclass[superscriptaddress,twocolumn,showpacs,preprintnumbers,amsmath,amssymb,prl,groupedaddress]{revtex4-1}
\fi
\usepackage[table]{xcolor}
\usepackage{slashed}
\usepackage{cancel}

\usepackage{multirow}
\usepackage[hidelinks]{hyperref}

\newcolumntype{s}{>{\columncolor[HTML]{AAACED}} p{3cm}}

\usepackage{amssymb}
\usepackage{epsfig,amsmath,graphics}
\usepackage{epstopdf}
\usepackage{graphicx}
\usepackage{color}
\usepackage{gensymb}

\usepackage[utf8x]{inputenc}
\newcommand{\ket}[1]{\ensuremath{\left|#1\right>}}

\renewcommand{\vec}{\mathbf}

\usepackage{natbib}
\usepackage{graphicx}

\begin{document}

\title{Ytterbium atom interferometry for dark matter searches}

\author{Yifan Zhou}
\affiliation{Department of Physics and Astronomy, The Johns Hopkins University, Baltimore, Maryland 21218}
\author{Rowan Ranson}
\affiliation{Department of Physics and Astronomy, The Johns Hopkins University, Baltimore, Maryland 21218}
\author{Michalis Panagiotou}
\affiliation{Department of Physics and Astronomy, The Johns Hopkins University, Baltimore, Maryland 21218}
\author{Chris Overstreet}
\email{c.overstreet@jhu.edu}
\affiliation{Department of Physics and Astronomy, The Johns Hopkins University, Baltimore, Maryland 21218}

\begin{abstract}
We analyze the projected sensitivity of a laboratory-scale ytterbium atom interferometer to scalar, vector, and pseudoscalar dark matter signals.  A frequency ratio measurement between two transitions in $^{171}$Yb enables a search for variations of the fine-structure constant that could surpass existing limits by a factor of 100 in the mass range $10^{-22}$ eV to $10^{-16}$ eV.  Differential accelerometry between ytterbium isotopes yields projected sensitivities to scalar and vector dark matter couplings that are stronger than the limits set by the MICROSCOPE equivalence principle test, and an analogous measurement in the MAGIS-100 long-baseline interferometer would be more sensitive than previous bounds by factors of 10 or more.  A search for anomalous spin torque in MAGIS-100 is projected to reach similar sensitivity to atomic magnetometry experiments.  We discuss strategies for mitigating the main systematic effects in each measurement.  These results indicate that improved dark matter searches with ytterbium atom interferometry are technically feasible.

\end{abstract}
\maketitle

\section{I.  Introduction}  

The nature of dark matter is one of the most pressing open questions in fundamental physics.  Many astronomical observations point to the existence of dark matter \cite{Freese2009}, and cosmological models indicate that its energy density near the Earth should be about $0.4$ GeV/cm$^3$ \cite{JacksonKimball2023}.  Beyond that, the properties of the dark matter remain unknown.  In particular, the mass of a dark matter particle could be anywhere between $10^{-22}$ eV and $10^{28}$ eV.  The lower bound on the dark matter particle mass is set by assuming that the dark matter's de Broglie wavelength is no larger than a dwarf galaxy, while the upper bound is set by the Planck scale.  

Phenomenologically, the signal in a dark matter direct detection experiment changes qualitatively at a dark matter particle mass of $\sim 1$ eV.  If the dark matter particle mass is above $1$ eV, the phase space density of the dark matter near the Earth is less than one, and the dark matter is expected to exhibit particle-like behavior (e.g., scattering off of ordinary matter).  Several experiments \cite{Alkhatib2021, Abdallah2015} have searched for dark matter in this regime.  If the dark matter particle mass is below $1$ eV, however, the phase space density of the dark matter near the Earth is greater than one.  In this case, the dark matter must be a boson, and its behavior is analogous to that of a classical field:  all of its lowest-order couplings to the Standard Model yield signals that oscillate at the Compton frequency of the dark matter \cite{Graham2016, Arvanitaki2015}.  Experiments searching for this ``ultralight'' dark matter include optical clocks \cite{Filzinger2023, Sherrill2023}, microwave cavity experiments \cite{Bartram2021}, nuclear magnetic resonance experiments \cite{Aybas2021, Wu2019}, and atomic magnetometers \cite{Bloch2020, Lee2023}, among others.  Ultralight dark matter would also give rise to new static forces between ordinary matter \cite{footnote1}.  Such forces have been constrained by equivalence principle tests \cite{Touboul2022, Schlamminger2008}. 

In this work, we consider the potential physics reach of dark matter detection experiments based on atom interferometry.  In a light-pulse atom interferometer \cite{Hogan2009}, ultracold atoms are split by atom-light interactions into a superposition of external states, which are then recombined and interfered.  Depending on the interferometer geometry, the phase of an atom interferometer can be sensitive to inertial forces \cite{Kasevich1991}, recoil velocity \cite{Borde1989}, and/or the energy difference between internal states \cite{Dimopoulos2008}.  Atom interferometers have been used to test the equivalence principle at a relative accuracy of about $10^{-12}$ \cite{Asenbaum2020a}, measure the fine-structure constant \cite{Morel2020,Parker2018}, and observe gravitational phase shifts in nonlocal quantum systems \cite{Overstreet2022,Asenbaum2017}.  In addition, several gravitational wave detectors based on atom interferometry \cite{Abe2021, Badurina2020, Canuel2018} are under construction. 

High-precision atom interferometers typically have a measurement time on the order of $1$ s and a cycle time of a few seconds; thus, atom interferometers are naturally sensitive to dark matter with Compton frequency $\lesssim 1$ Hz.  With a measurement campaign of about one year, an atom interferometry experiment can search for oscillating dark matter signals over eight orders of magnitude in the dark matter particle mass (from $10^{-8}$ Hz to $1$ Hz, or $10^{-22}$ eV to $10^{-14}$ eV). The possible signals from ultralight dark matter depend on its spin and parity.  A spin-zero or spin-one dark matter field can produce six qualitatively different signals at lowest order \cite{Graham2016}.  Of these, atom interferometers can be sensitive to three:  variations of fundamental constants, accelerations, and spin torques.  In addition, atom interferometers can search for static forces induced by dark matter. 

One of the main considerations for an atom-interferometric dark matter detection experiment is the choice of atomic species.  Ideally, the experiment should use an atom that provides high sensitivity to dark matter signals while minimizing systematic effects.  Here we consider atom interferometry with ytterbium isotopes.  Ytterbium offers the highest sensitivity of any neutral atom to variations of the fine-structure constant in one of its excited states \cite{Safronova2018, Dzuba2018}, possesses multiple isotopes that can be cooled simultaneously for a differential acceleration measurement \cite{Kitagawa2008}, and has an isotope with nuclear spin $1/2$ that can be used to detect spin torque.  In addition, the alkaline-earth-like electronic structure of ytterbium provides magnetic insensitivity in the $^1S_0$ electronic ground state, and its multiple narrow-linewidth transitions facilitate precise measurements of transition frequencies.  Atom interferometry with ytterbium has previously been demonstrated \cite{Gochnauer2021} and has previously been proposed for tests of fundamental physics \cite{Hartwig2015}.  

A laboratory-scale ytterbium atomic fountain experiment is currently under construction at Johns Hopkins University.  In this work, we calculate the projected sensitivity of this experiment to dark matter couplings.  We find that atom interferometry utilizing two clock transitions in $^{171}$Yb can provide a hundredfold improvement in searches for variations of the fine-structure constant.  These transitions have previously been identified for an optical-clock-based search for scalar dark matter \cite{Safronova2018, Dzuba2018}.  We also project that a static equivalence principle test in the JHU apparatus can be more sensitive than the MICROSCOPE experiment to scalar and vector dark matter couplings.  Finally, we show that although a search for spin torque in laboratory-scale atom interferometers is unlikely to reach the limits set by atomic magnetometry experiments, an analogous search in a long-baseline interferometer would have comparable sensitivity. 

The remainder of this paper is organized as follows.  The apparatus is described in Section~\hyperref[sec:Apparatus]{II}, and its sensitivities to scalar, vector, and pseudoscalar dark matter are discussed in Sections~\hyperref[sec:Scalar]{III.A}, \hyperref[sec:Vector]{III.B}, and \hyperref[sec:Axion]{III.C}, respectively. Section \hyperref[sec:Systematics]{IV} describes the main systematic effects in these measurements and how they will be controlled.  Section \hyperref[sec:Discussion]{V} compares atom-interferometric dark matter searches to other experiments and discusses the prospects for future sensitivity improvements.

\section{II.  Apparatus description}
\label{sec:Apparatus}

The JHU experimental apparatus will consist of a source of ultracold ytterbium and an atomic fountain in which the atoms can freely fall during interferometry sequences.  Clouds of ultracold ytterbium will be produced in a 3D magneto-optical trap (MOT) that is loaded by a commercial Zeeman slower/2D MOT.  The 3D MOT will utilize core-shell techniques \cite{Lee2015} to increase atom number and phase space density.  The atoms will be evaporatively cooled in an optical dipole trap and then launched into a magnetically shielded atomic fountain by an optical lattice.  The height of the magnetically shielded region will be $2.5$ m, enabling interferometry over a free-fall distance of $2$ m.  After the interferometry sequence is complete, the interferometer phase will be measured via fluorescence detection of the number of atoms in each output port.     

The apparatus will be capable of driving several electronic transitions for atom interferometry.  These include Bragg transitions on the $^1S_0 \leftrightarrow {}^1P_1$ transition at $399$ nm and the $^1S_0 \leftrightarrow {}^3P_1$ transition at $556$ nm, as well as single-photon transitions at $556$ nm, $578$ nm ($^1S_0 \leftrightarrow {}^3P_0$), and $431$ nm [$^1S_0 \leftrightarrow 4\text{f}^{13}6\text{s}^{2}5\text{d}\,(J = 2)$].  We also consider the possibility of driving the transitions at $578$ nm and $431$~nm with a Doppler-free two-photon process, which would simplify some of the interferometer geometries considered in Section~\hyperref[sec:Sensitivity]{III}. 

For the purpose of generating estimated sensitivities to dark matter, we assume that $10^{5}$ atoms will participate in each measurement.  We also assume that the coherence time of each measurement will be $1.3$ s, limited by the available free fall time, and that the experimental cycle time will be $10$ s.  These atom numbers and cycle times have previously been achieved in high-precision atom interferometers \cite{Asenbaum2020a}.  Finally, we assume that each measurement campaign will have a duration of one year.

\section{III.  Dark matter detection with Yb atom interferometry}
\label{sec:Sensitivity}

Atom interferometers naturally detect energy level shifts and accelerations.  With the appropriate geometry, an atom interferometer can be sensitive to scalar, vector, or pseudoscalar (axion-like) couplings to ordinary matter.  In this section, we describe interferometer geometries suitable for detecting each of these dark matter candidates and calculate the projected sensitivity of the JHU apparatus to the associated coupling.  For each estimate, we assume shot-noise-limited statistics and a one-year measurement campaign.  The leading systematic effects for each measurement are discussed in Section~\hyperref[sec:Systematics]{IV}. 

Searches for oscillating signals produced by dark matter rely on assumptions about the distribution of dark matter in our galaxy.  Following Ref. \cite{Graham2016}, we assume that the dark matter is distributed according to a standard halo model with energy density $\rho_\text{DM} = 0.4$ GeV/cm$^3$.  We also assume that a single dark matter species comprises this energy density.  The amplitude of the dark matter field is then proportional to $\sqrt{2 \rho_\text{DM}}/(m_\text{DM} c)$, where $m_\text{DM}$ is the dark matter particle mass and $c$ is the speed of light.  The oscillation frequency of the dark matter field is set by its Compton frequency $m_\text{DM} c^2/\hbar$, where $\hbar$ is the reduced Planck's constant.  Finally, we assume that the magnitude of the dark matter velocity is approximately equal to the virial velocity, $10^{-3}\, c$, and we model the unknown direction of the dark matter velocity (and polarization, for vector dark matter) with a uniformly distributed random variable.  The dark matter velocity spread implies a frequency spread of one part in $10^6$, which limits the maximum integration time of a coherent measurement to $10^6$ oscillations.  These assumptions are incorporated into the projected sensitivites to oscillating signals in each of the following sections.  In addition to oscillating signals, a dark matter particle can give rise to new static forces, and limits on dark matter couplings derived from static searches do not depend on any assumptions about the galactic dark matter distribution.

\subsection{A.  Scalar dark matter}
\label{sec:Scalar}

Scalar particles are appealing as dark matter candidates because they have a natural production mechanism in the early universe \cite{Dine1983, Preskill1983} and because many extensions of the Standard Model include new scalars \cite{Damour2010}.  At lowest order, scalar particles can interact with ordinary matter through dilaton couplings or through the Higgs portal \cite{Graham2016}.  Here we consider the dilaton coupling of the scalar field to the electromagnetic field, which is represented by the Lagrangian density term

\begin{equation} \label{eq:ScalarEMcoupling}
\mathcal{L} \supset \frac{d_e}{4 \mu_0} \varphi F_{\mu \nu}F^{\mu \nu}
\end{equation}
where $\mu_0$ is the vacuum permeability, $\varphi$ is the scalar field, $F$ is the electromagnetic field tensor, and $d_e$ is a dimensionless coupling constant.  
% To account for the energy density $\rho_\text{DM} \approx 0.4 $ GeV/cm$^3$ of dark matter near the Earth, 
The amplitude of the scalar field is given by
\begin{equation} \label{Eq:scalarAmplitude}
\varphi_0 = \sqrt{\frac{4 \pi G \hbar^2}{c^4}} \frac{\sqrt{2 \rho_\text{DM}}}{m_\text{DM} c}
\end{equation}
where $G$ is the gravitational constant.  As discussed in Ref. \cite{Arvanitaki2015}, this interaction leads to an apparent variation of the fine-structure constant $\alpha$, which is given (at lowest order in $d_e$) by 
\begin{equation} \label{Eq:alphaT}
    \alpha(t) = \alpha_0 \left(1 + d_e \varphi_0 \cos \left(\frac{m_\text{DM} c^2}{\hbar} t \right) \right)  
\end{equation} 
where $\alpha_0$ is the unperturbed value.  Experiments that can detect variations of $\alpha$ can therefore search for scalar dark matter.

Atom interferometers are sensitive to the value of the fine-structure constant through its influence on atomic transition frequencies.  The dependence of a transition frequency $\omega$ on variations of $\alpha$ can be parameterized by a constant $\Delta q$ as follows \cite{Safronova2018, Dzuba2018}: 
\begin{equation}
    \omega(\alpha) = \omega(\alpha_0) + \frac{\Delta q}{\hbar} \left[ \left(\frac{\alpha(t)}{\alpha_0}\right)^2 - 1  \right].
\end{equation}
Note that $\Delta q$ describes the relativistic corrections to the transition frequency \cite{Dzuba2018}. A frequency ratio measurement between two transitions is sensitive to variations of $\alpha$ as long as the value of $\Delta q$ differs between them, and transitions with large values of $\Delta q$ have the highest discovery potential.  

\begin{figure}
    \centering
    \includegraphics[width=0.40\textwidth]{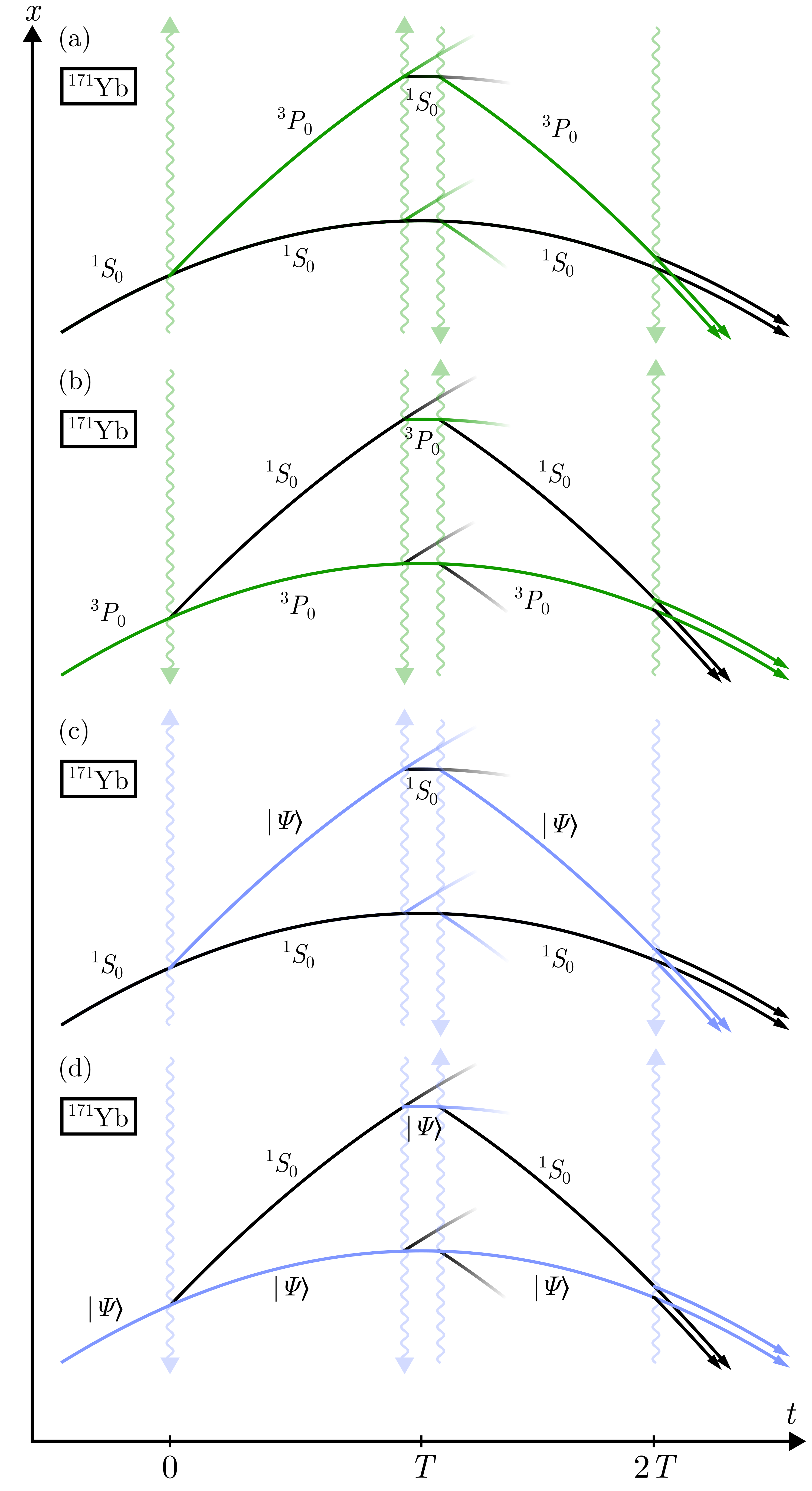}
    \caption{Interferometer geometry for searching for variations of the fine-structure constant.  Light pulses at times $t = 0$, $t = T$, and $t = 2T$ drive single-photon transitions between the $^1S_0$ electronic ground state and one of the excited states, $^3P_0$ or $\ket{\Psi} = [\text{Xe}]{4\text{f}^{13} 6\text{s}^2 5\text{d}}\,(J = 2)$. The differential phase shifts between pairs of Ramsey-Bord\'{e} interferometers allow each transition frequency to be measured independently, and the frequency ratio is sensitive to variations of the fine-structure constant but insensitive to laser frequency drift.  To reduce systematic effects, the four interferometers are spatially overlapped; interferometers are spaced vertically in the diagram for clarity.}
    
    \label{fig:Alpha_time}
\end{figure}

\begin{figure}
    \centering
    \includegraphics[width=0.4\textwidth]{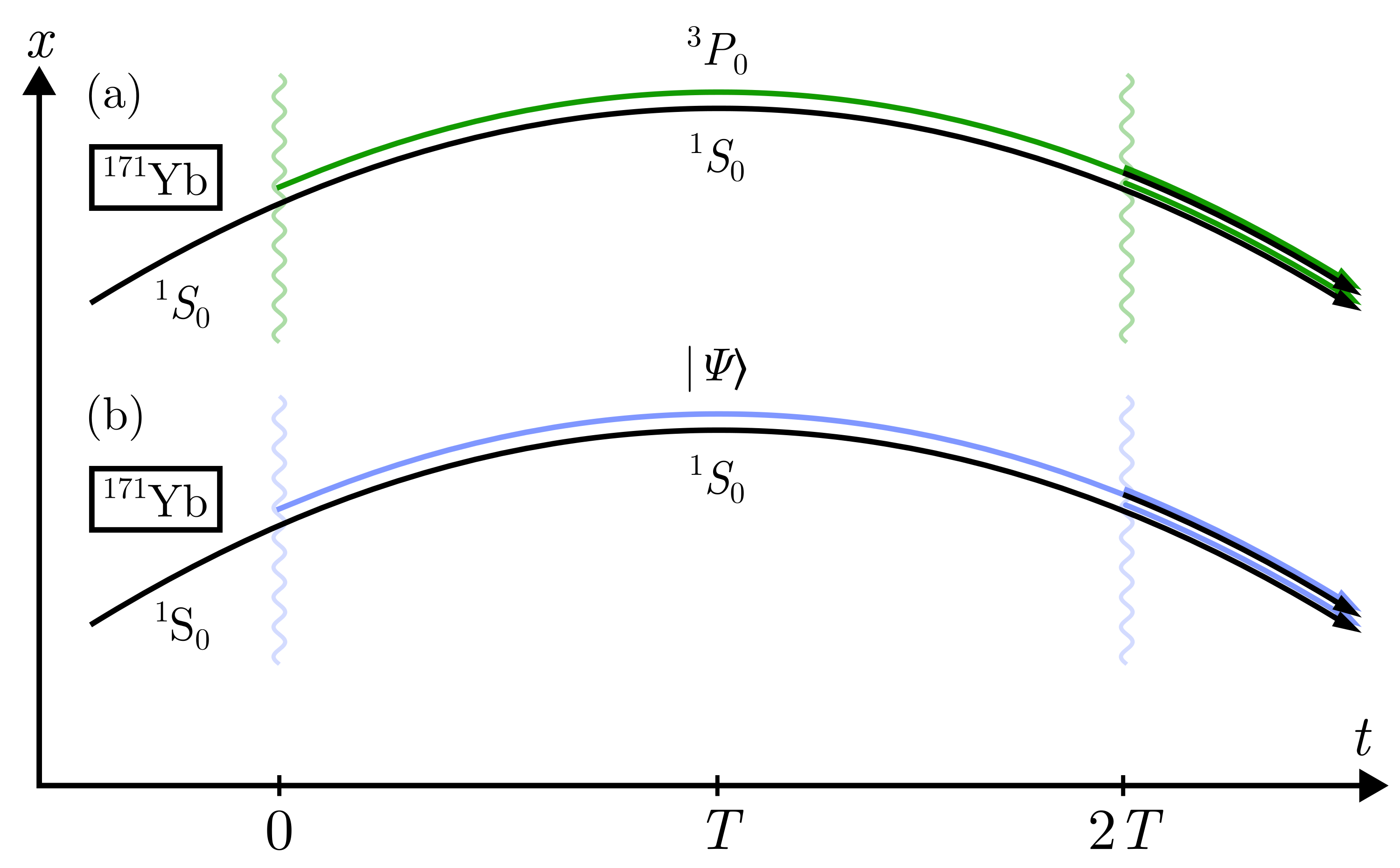}
    \caption{Alternative interferometer geometry for searching for variations of the fine-structure constant. Doppler-free two-photon transitions at time $t = 0$ create superpositions of the $^1S_0$ electronic ground state and one of the excited states, $^3P_0$ or $\ket{\Psi} =$ [Xe]4f$^{13}$6s$^{2}$5d $(J = 2)$. The phase of each interferometer is sensitive to the transition frequency between the internal states of its arms. To reduce systematic effects, the two interferometers are spatially overlapped; interferometers are spaced vertically in the diagram for clarity.}
    
    \label{fig:Two-photon_alpha_time}
\end{figure}

We propose to use a $^{171}$Yb atom interferometer to measure the frequency ratio between the $^1S_0 \leftrightarrow {}^3P_0$ transition at $578$ nm and the $^1S_0 \leftrightarrow 4\text{f}^{13}6\text{s}^{2}5\text{d}\,(J = 2)$ transition at $431$ nm.  The energy of the $4\text{f}^{13}6\text{s}^{2}5\text{d}\,(J = 2)$ state has the highest sensitivity among neutral-atom states to variations of $\alpha$ \cite{Safronova2018}.  An interferometer geometry for this frequency comparison is depicted in Fig.~\ref{fig:Alpha_time}.  In each of the four Ramsey-Bord\'{e} interferometers, a single-photon transition creates a superposition of internal states, which are interfered to produce an interferometer phase  of the form \cite{Hogan2009, Dimopoulos2008}
\begin{equation}
\phi = 2 (\omega - \omega_L) T + \frac{\hbar \omega_L}{m c^2}(2 \omega - \omega_L) T + \frac{\omega}{c} g T^2 + \cdots. 
\end{equation}
Here $\omega$ is the transition frequency, $\omega_L$ is the laser frequency, $T$ is the interferometer time, $m$ is the atomic mass, and $g$ is the magnitude of the laboratory acceleration relative to a freely falling geodesic.  The first term is the desired signal, while the second and third terms represent systematic effects due to the recoil velocity and the laboratory acceleration, respectively \cite{footnote3}.  To suppress these effects, we consider the differential phase between pairs of interferometers [(a) - (b) and (c) - (d)], which has the form
\begin{equation} \label{Eq:scalarAC}
\Delta \phi = 2 \int_0^{2T} \omega(t)\, dt - 4\, \omega_L T + \cdots 
\end{equation}
and is insensitive to the phase shifts arising from the recoil velocity and the laboratory acceleration because they are common to both interferometers. In Eq.~\ref{Eq:scalarAC}, we have explicitly included the time dependence of $\omega$.  For most of the dark matter parameter space we consider, the oscillation period of $\omega$ is much longer than the interferometer time, and $\omega$ is approximately constant during each measurement.  For higher dark matter masses, the oscillation of $\omega$ during a single measurement becomes significant and leads to loss of sensitivity.  In addition, we note that this phase shift has the same form as that of an optical lattice clock interrogated with a Ramsey sequence.  As in an optical clock, the population difference between the two output states will be a function of the frequency difference between the atomic transition and the laser.  

Both of the single-photon transitions depicted in Fig.~\ref{fig:Alpha_time} can be driven with technically feasible laser systems.  For the $^1S_0 \leftrightarrow {}^3P_0$ transition at $578$ nm, which is weakly allowed in $^{171}$Yb due to the nonzero nuclear spin, a laser intensity of $1$~W/cm$^2$ is sufficient to obtain a Rabi frequency of $40$~kHz. 
 The $^1S_0 \leftrightarrow 4\text{f}^{13}6\text{s}^{2}5\text{d}\,(J = 2)$ transition at $431$ nm is an M2 transition with Rabi frequency $17$~Hz $\cdot\, \sqrt{I}$/(mW/cm$^2$)$^{1/2}$ \cite{Dzuba2018}.  Thus, a Watt-class laser with roughly $1$~cm beam size would enable $1$~kHz Rabi frequency.  The lifetime of the $4\text{f}^{13}6\text{s}^{2}5\text{d}\,(J = 2)$ state has been estimated to be between $60$~s and $200$~s, limited by decay to the triplet $P$ manifold \cite{Safronova2018,Dzuba2018}.   

An alternative geometry for measuring this frequency ratio is shown in Fig.~\ref{fig:Two-photon_alpha_time}, where each beamsplitter is implemented by a Doppler-free two-photon transition driven by counterpropagating laser beams \cite{Alden2014}.  Compared to the geometry in Fig.~\ref{fig:Alpha_time}, Doppler-free transitions naturally suppress the recoil shift, thereby reducing the number of interferometers required from four to two, and prevent the atom loss into undesired momentum states that is intrinsic to Ramsey-Bord\'e interferometers.  However, Doppler-free two-photon transitions require higher laser power than single-photon transitions for a given Rabi frequency. 

\begin{figure}
    \centering
    \includegraphics[width=0.40\textwidth]{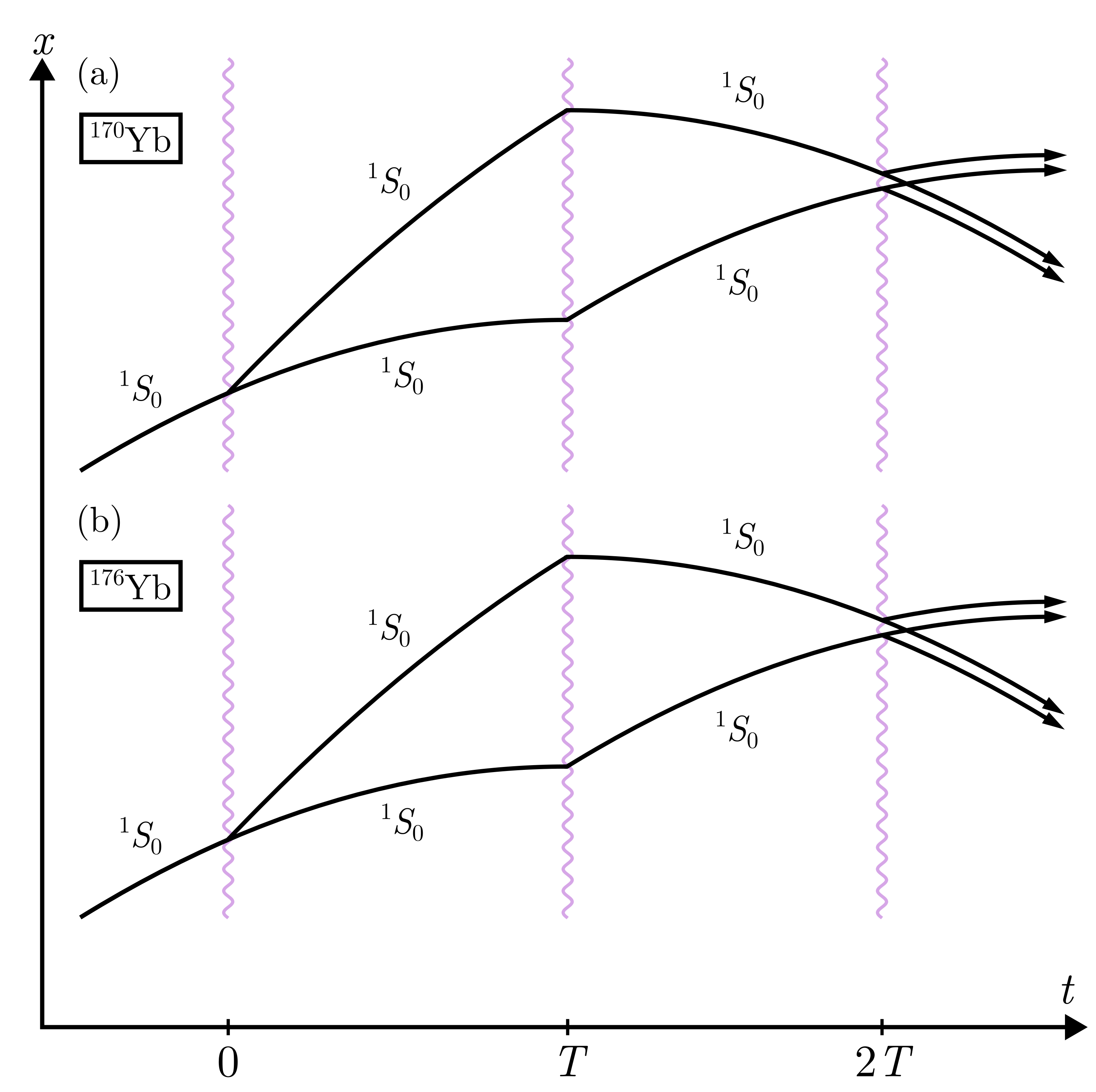}
    \caption{Interferometer geometry for detecting differential acceleration induced by dark matter.  Each beamsplitter and mirror pulse in the Mach-Zehnder interferometers is implemented with Bragg transitions, e.g. using $399$ nm or $556$~nm light, so that each interferometer arm remains in the $^1S_0$ ground state.  The differential phase shift is sensitive to the acceleration difference between isotopes.  The use of a common laser to address both isotopes suppresses phase shifts due to vibrations of the retroreflection mirror and laser phase noise. To reduce systematic effects, the two interferometers are spatially overlapped; interferometers are spaced vertically in the diagram for clarity.}
    
    \label{fig:Isotope_gradiometer}
\end{figure}

In addition to an oscillating variation of fundamental constants, the existence of scalar dark matter would also give rise to a time-independent Yukawa potential between Standard Model particles with the following form \cite{Damour2010}:
\begin{equation}\label{eq:Yukawa Potential}
    V_\text{scalar}(r)=-\frac{G m_1 m_2}{r}\, d_e^2\, Q_{1}\, Q_{2}\, e^{-r/\lambda}.
\end{equation}
Here $r$ is the distance between particles, $m_i$ and $Q_i$ are the mass and the dilaton charge of particle $i$, respectively, and $\lambda$ is the Compton wavelength of the dark matter particle.  Note that the coupling constant $d_e$ in this expression is the same quantity as in Eq.~\ref{eq:ScalarEMcoupling}. Since the dilaton charge of an atom is a function of its mass number and atomic number (see Appendix~\hyperref[app:charges]{A.I}), this potential leads to a composition-dependent static force   
\begin{equation}\label{eq:ScalarDCForce}
    \vec{F}_\text{scalar}^\text{DC} = \frac{G m_1 m_2}{r^2}\, d_e^2\, Q_{1}\, Q_{2}\, \left(1 + \frac{r}{\lambda} \right)e^{-r/\lambda}\, \hat{\mathbf{r}}
\end{equation}
where $\hat{\mathbf{r}}$ is the unit vector pointing from one particle to the other.  An ultralight scalar particle can therefore be detected by measuring the differential acceleration between two atomic species in the Yukawa potential sourced by the Earth---in other words, by performing an equivalence principle test.  The interferometer geometry for such a test is shown in Fig.~\ref{fig:Isotope_gradiometer}.  In this Mach-Zehnder gradiometric configuration, the differential phase is given by \cite{Hogan2009}
\begin{equation} \label{Eq:MachZehnder}
    \Delta \phi = n\, k\, \Delta g\, T^2 + \cdots
\end{equation}
where $n$ is the number of photon recoils in the initial beamsplitter, $k$ is the magnitude of the laser wavevector, and $\Delta g$ is the acceleration difference between isotopes projected onto the interferometer axis.  For this measurement, we consider a comparison between $^{170}$Yb and $^{176}$Yb, which are both bosons and have favorable scattering lengths \cite{Kitagawa2008} that allow simultaneous evaporative cooling.  The beamsplitters can be implemented by Bragg transitions on the electric dipole transition at $399$~nm or Bragg transitions on the intercombination transition at $556$~nm.  Either sequential two-photon transitions \cite{Asenbaum2020a} or multi-photon transitions \cite{Parker2018} could be used to impart momentum to the atoms.  We assume $n = 500$, allowing a maximum wave packet separation of $2$ m at time $T$.  Beamsplitters with hundreds of photon recoils have previously been demonstrated \cite{Wilkason2022, Beguin2023}.  With these parameters, the JHU apparatus is projected to reach a sensitivity to the E\"{o}tv\"{o}s parameter $\eta \equiv \Delta g/g = 2 \times 10^{-16}$ after a one-year measurement campaign.

\begin{figure}
    \centering
    \includegraphics[width=0.45\textwidth]{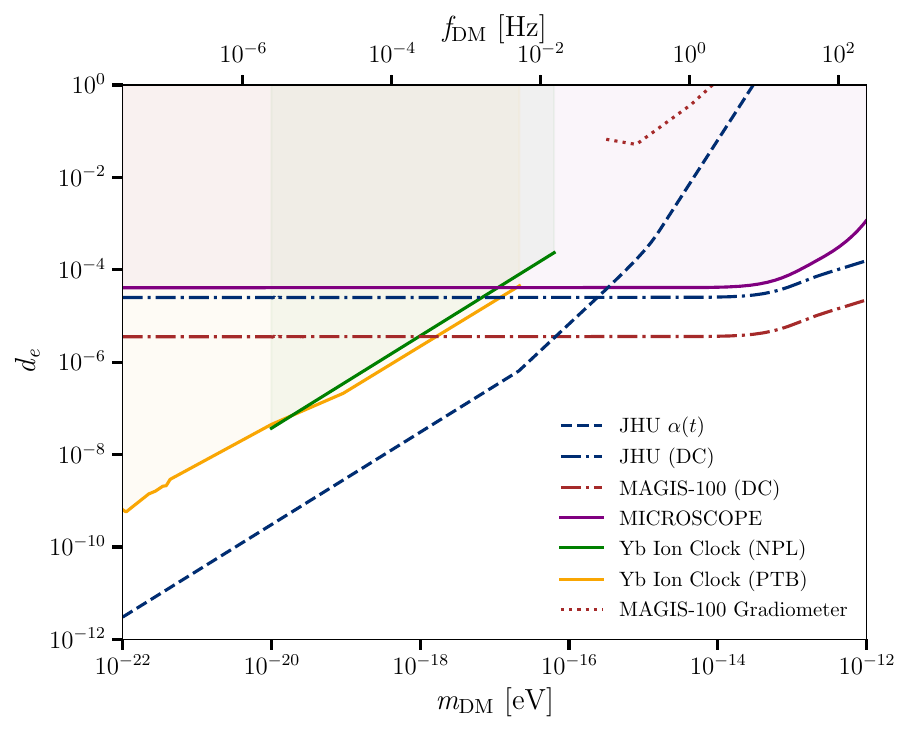}
    \caption{Constraint plot and projected sensitivities to scalar dark matter coupled to the electromagnetic field tensor.  JHU $\alpha(t)$: projected sensitivity of spectroscopic search in JHU apparatus with interferometer geometry shown in Fig. 1.  JHU (DC):  projected sensitivity of static equivalence principle test in JHU apparatus with interferometer geometry shown in Fig. 3.  MAGIS-100 (DC):  projected sensitivity of static equivalence principle test with Yb isotopes in MAGIS-100 long-baseline interferometer.  MAGIS-100 Gradiometer:  projected sensitivity of gradiometric measurement proposed in Ref.~\cite{Abe2021}, assuming the same phase resolution and interferometer parameters as other projections.  Also shown are existing constraints from the MICROSCOPE experiment \cite{Touboul2022} (purple curve) and Yb ion clock experiments at NPL \cite{Sherrill2023} (green curve) and PTB \cite{Filzinger2023} (yellow curve).  Shaded regions are excluded by existing constraints.  All curves are 95\% confidence limits.}
    \label{fig:constraints_alpha_variation}
\end{figure}

\begin{figure}
    \centering
    \includegraphics[width=0.45\textwidth]{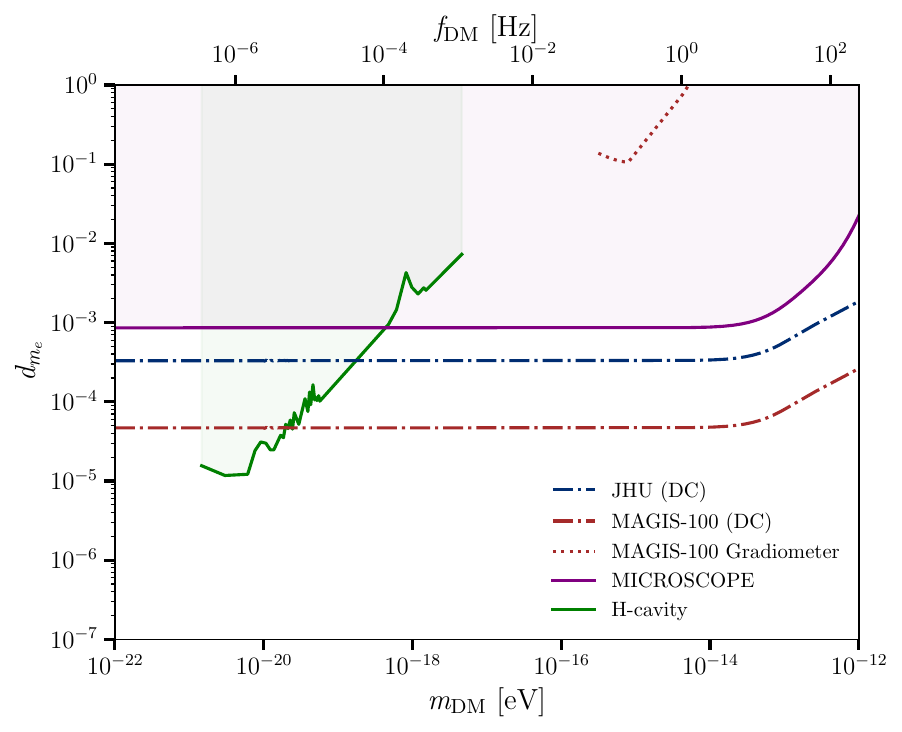}
    \caption{Constraint plot and projected sensitivities to scalar dark matter coupled to the electron mass.  JHU (DC):  projected sensitivity of static equivalence principle test in JHU apparatus with interferometer geometry shown in Fig. 3.  MAGIS-100 (DC):  projected sensitivity of static equivalence principle test with Yb isotopes in MAGIS-100 long-baseline interferometer.  MAGIS-100 Gradiometer:  projected sensitivity of gradiometric measurement proposed in Ref.~\cite{Abe2021}, assuming the same phase resolution and interferometer parameters as other projections.  Also shown are existing constraints from the MICROSCOPE experiment \cite{Touboul2022} (purple curve) and a hydrogen maser-optical cavity comparison \cite{Kennedy2020} (green curve).  Shaded regions are excluded by existing constraints.  All curves are 95\% confidence limits. }
    
    \label{fig:constraints_d_m_e}
\end{figure}

The projected sensitivities of these measurements to the scalar dark matter coupling constant $d_e$ are shown in Fig.~\ref{fig:constraints_alpha_variation} as a function of the dark matter particle mass.  The projections are compared to existing constraints from ytterbium ion clock experiments \cite{Filzinger2023, Sherrill2023} and the MICROSCOPE space-based equivalence principle test \cite{Touboul2022}.  We calculate the projected sensitivity to variations of $\alpha$ both analytically and by taking the discrete Fourier transform of simulated data.  In the JHU experiment, the Lomb-Scargle periodogram \cite{VanderPlas2018} will be used instead of the discrete Fourier transform to optimally account for variations in cycle time and experimental dead time.  The projections take into account the stochastic amplitude of the dark matter field, which diminishes sensitivity to oscillating signals by a factor of about 3 when the measurement campaign is shorter than the dark matter coherence time \cite{Centers2021}.  We conservatively apply this sensitivity reduction across the entire parameter space.  A more sophisticated analysis, following Ref. \cite{Centers2021}, would improve the projected sensitivity at frequencies above $10^{-1}$ Hz by a factor of order one.    

At low dark matter masses, the projected sensitivity of the search for variations of $\alpha$ scales like $1/m_\text{DM}$ (see Eqs.~\ref{Eq:scalarAmplitude} and \ref{Eq:alphaT}), while the projected sensitivity of the equivalence principle test becomes independent of $m_\text{DM}$ (see Eq.~\ref{eq:ScalarDCForce}). At higher dark matter masses, the search for variations of $\alpha$ loses sensitivity due to signal averaging over the $\sim 1$~s coherence time of each experimental run. To avoid the loss of sensitivity when $m_\text{DM}c^2/\hbar \cdot T = n \pi$ for integer $n$, the interferometer time $T$ can be varied, as described in Ref.~\cite{Badurina2022}. When $m_\text{DM}c^2/\hbar$ exceeds the Nyquist frequency, searches for dark matter can be performed using the techniques introduced in Ref.~\cite{Badurina2023}, which we do not consider here.  In contrast, the equivalence principle test loses sensitivity at higher dark matter masses due to the exponential decay term in Eq.~\ref{eq:ScalarDCForce}, which effectively limits the quantity of material in the Earth that sources the static force.  The change in slope at $m_\text{DM} \sim 10^{-13}$~eV arises from our model of the Earth, which takes into account the composition difference between the core and the mantle (see Appendix~\hyperref[app:earth]{A.II}). 

We also plot the projected sensitivity of an equivalence principle test performed between $^{170}$Yb and $^{176}$Yb in the MAGIS-100 long-baseline interferometer at Fermilab \cite{Abe2021}.  For this projection, we use the interferometer geometry in Fig.~\ref{fig:Isotope_gradiometer} and the same experimental parameters as in the JHU apparatus (atom number, photon recoils, etc.), except that the interferometer time is increased from $T = 0.64$ s to $T = 4.5$ s.  We note that this measurement in MAGIS-100 would reach an E\"{o}tv\"{o}s parameter sensitivity of $\eta = 5 \times 10^{-18}$.  

The JHU atom-interferometric search for variations of the fine-structure constant is projected to improve on existing experiments by two orders of magnitude in the mass range $10^{-22}$ eV to $10^{-16}$ eV.  The ytterbium equivalence principle test performed in the JHU apparatus is projected to reach a similar sensitivity to the MICROSCOPE experiment, while the MAGIS-100 equivalence principle test would be a factor of 10 more sensitive than existing bounds from $10^{-17}$ eV to $10^{-12}$ eV.  

An equivalence principle test is also sensitive to other possible couplings of scalar dark matter to the Standard Model.  For example, Fig.~\ref{fig:constraints_d_m_e} shows the projected sensitivities of the JHU and MAGIS-100 experiments to the coupling of a scalar field to the electron mass, parameterized by $d_{m_e}$. This coupling would induce a static force with the same form as Eq.~\ref{eq:ScalarDCForce}, but with $d_e \rightarrow d_{m_e}$ and the appropriate dilaton charges for each particle (see Appendix~\hyperref[app:charges]{A.I}).  For this coupling, the JHU and MAGIS-100 measurements are projected to be more sensitive than MICROSCOPE by factors of $3$ and $20$, respectively.   

We note that a gradiometer utilizing a single optical transition has also been proposed to search for scalar dark matter \cite{Arvanitaki2018,Abe2021}.  The projected sensitivity of such an experiment in MAGIS-100 to $d_e$ and $d_{m_e}$ is shown in Figs. 4 and 5, respectively. A ytterbium gradiometer would have similar projected sensitivity to a strontium gradiometer for these searches.  Nevertheless, we project that the gradiometric operating mode would be less sensitive than the bounds set by the MICROSCOPE experiment throughout the mass range.  The discrepancy between our projection and previous work \cite{Arvanitaki2018,Abe2021} is explained by our more conservative estimate of the attainable phase resolution \cite{footnote2}.  In addition, we note that gradiometric searches for dark matter at low frequencies are severely constrained by gravity gradient noise, which becomes challenging to characterize at frequencies below $1$~Hz \cite{Badurina2023b}.  Since the differential phase response to gravity gradients scales linearly with the spatial separation between two interferometers, the co-located interferometers in this proposal are much less susceptible to gravity gradient noise.

\subsection{B.  Vector dark matter}
\label{sec:Vector}

Next, we consider the possibility that the dark matter consists of a vector particle.  Vector particles have a natural production mechanism in the early universe through inflationary fluctuations \cite{Graham2016a} and are thus cosmologically well-motivated as dark matter candidates.  Vector dark matter can exert forces on ordinary matter through the minimal coupling to fermions.  Here we consider a vector coupling to the charge $B - L$, the baryon number minus the lepton number, with coupling constant $g_{B - L}$.  For neutral atoms, this charge is equal to $A - Z$, the neutron number. 

The interaction between atoms and the galactic dark matter field leads to a time-varying force that oscillates at the Compton frequency of the dark matter particle \cite{Graham2016,Shaw2022},
\begin{equation}
    \vec{F}^\text{AC}_\text{vector} = g_{B-L}\,\sqrt{2\rho_{\text{DM}}}\,(A - Z)\,\cos{\left({\frac{m_{\text{DM}}c^2}{\hbar}t}\right)}\,\hat{\vec{e}},
\end{equation}
where $\hat{\vec{e}}$ points in the polarization direction of the dark matter field.
In the presence of a time-varying acceleration $g(t)$, the phase response of a Mach-Zehnder interferometer is given by \cite{Overstreet2021}
\begin{equation}
    \phi = n k\, \left[\int_T^{2T} \int_0^t g(t')\, dt'\, dt - \int_0^{T} \int_0^t g(t')\, dt'\, dt  \right] + \cdots.
\end{equation}
In addition, atoms can exert a force on one another by exchanging the vector particle.  This static force is given by 
\begin{equation}
    \vec{F}^{\text{DC}}_{\text{vector}} = \frac{g_{B-L}^2}{4\pi r^2}(A_1 - Z_1)(A_2 - Z_2)\left(1 + \frac{r}{\lambda} \right)e^{-r/\lambda}\,\hat{\vec{r}}
\end{equation}
where $A_i$ and $Z_i$ are the mass number and the atomic number of the $i$th particle, respectively, and $\hat{\vec{r}}$ is the unit vector pointing from one particle to the other.  
In an atom interferometer, both of these forces can be detected with a differential acceleration measurement between atomic species (Fig.~\ref{fig:Isotope_gradiometer}).  Detection of the static force sourced by the Earth requires an equivalence principle test in which systematic errors are controlled, while the time-oscillating force can in principle be detected even without accounting for DC systematic effects.   

\begin{figure}
    \centering
    \includegraphics[width=0.45\textwidth]{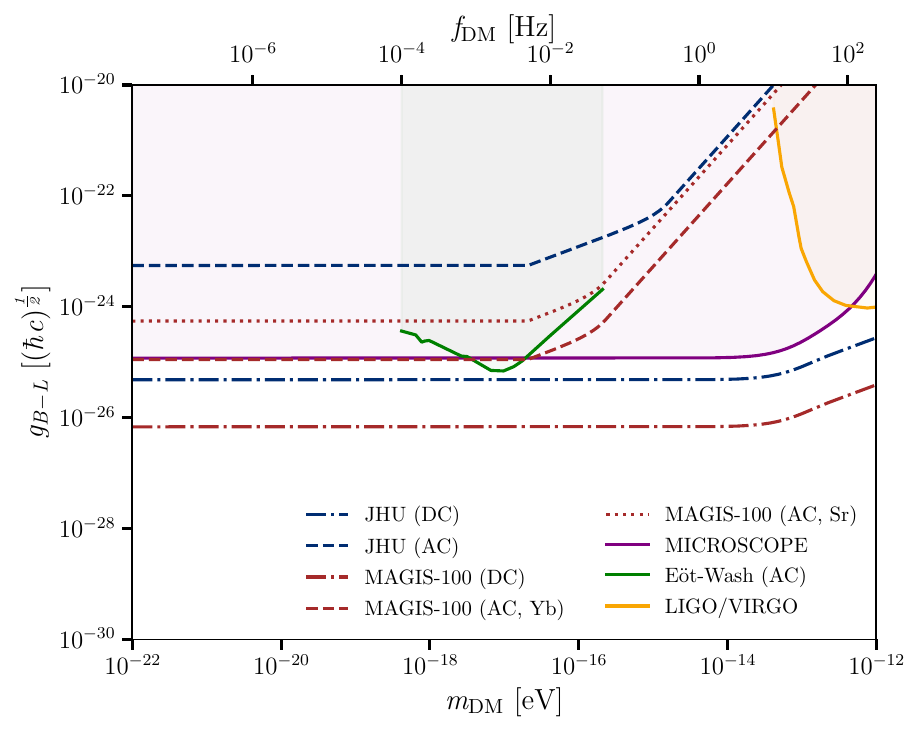}
    \caption{Constraint plot and projected sensitivities to vector dark matter coupled to $B - L$.  JHU (DC):  projected sensitivity of static equivalence principle test in JHU apparatus with interferometer geometry shown in Fig. 3.  JHU (AC):  projected sensitivity of search for time-dependent differential acceleration in JHU apparatus.  MAGIS-100 (DC):  projected sensitivity of static equivalence principle test with Yb isotopes in MAGIS-100 long-baseline interferometer.  MAGIS-100 (AC, Yb) and (AC, Sr):  projected sensitivities of searches for time-dependent differential acceleration in MAGIS-100 with Yb isotopes and Sr isotopes \cite{Abe2021}, respectively, assuming the same phase resolution and interferometer parameters as other projections.  Also shown are existing constraints from the MICROSCOPE experiment \cite{Touboul2022} (purple curve), the E\"{o}t-Wash experiment \cite{Shaw2022} (green curve), and the LIGO and VIRGO gravitational-wave detectors \cite{Abbott2022}.  Shaded regions are excluded by existing constraints.  All curves are 95\% confidence limits.}
    
    \label{fig:constraints_acceleration}
\end{figure}

Fig.~\ref{fig:constraints_acceleration} shows the projected sensitivity of atom-interferometric acceleration measurements to vector dark matter as a function of the dark matter particle mass.  For the JHU projections, we assume the same experimental parameters as in Section~\hyperref[sec:Scalar]{III.A} ($n = 500$, maximum wave packet separation $2$ m).  Likewise, the MAGIS-100 projections assume $n = 500$ and $T = 4.5$ s. 
Existing constraints from the MICROSCOPE \cite{Touboul2022}, E\"{o}t-Wash \cite{Shaw2022}, and LIGO/VIRGO experiments \cite{Abbott2022} are also shown.  The searches for oscillating dark matter take into account the stochastic amplitude of the dark matter field \cite{Centers2021} by reducing the projected sensitivity by a factor of 3 throughout the mass range.  These searches lose sensitivity at high frequencies when the period of the dark matter oscillation becomes shorter than the coherence time of a single measurement.  Sensitivity at high frequencies can be enhanced by resonant detection schemes \cite{Graham2016b}, which we do not consider here.  In addition, the sensitivities for oscillating signals are based on a single-peak search algorithm.  Using a multi-peak template \cite{Amaral2024} could provide higher sensitivity for some dark matter masses.  

The strongest searches for vector dark matter are expected to be derived from static equivalence principle tests.  In this operating mode, the JHU apparatus is projected to reach a similar sensitivity to the MICROSCOPE experiment, while an analogous test in MAGIS-100 would be more sensitive by a factor of 10.  Although the projected sensitivities of searches for time-oscillating forces are generally lower, we note that a search for oscillatory dark matter with ytterbium accelerometry in MAGIS-100 would reach a comparable sensitivity to MICROSCOPE and would be technically simpler than a static equivalence principle test.  Compared to previously proposed searches for time-oscillating forces with strontium \cite{Abe2021}, we find that our approach would be a factor of 5 more sensitive, assuming identical interferometer parameters, due to the larger isotopic difference in $B - L$ and the use of a higher-frequency transition for the beamsplitter pulses.

\subsection{C.  Pseudoscalar dark matter}
\label{sec:Axion}

Pseudoscalar particles (axion-like particles) are well-motivated dark matter candidates because they appear in theories that attempt to resolve other outstanding issues in fundamental physics, such as the strong CP problem \cite{Peccei1977} and the hierarchy problem \cite{Graham2015}.  Models inspired by string theory \cite{Arvanitaki2010} also predict the existence of low-mass axion-like particles.  Here we consider the detection of such particles by means of the spin torque that they exert on nucleons.  In the presence of a pseudoscalar field, a nuclear spin $\vec{s}$ experiences a Hamiltonian \cite{Graham2018}
\begin{equation}
    H_a = -g_{\text{aNN}}\, \vec{\nabla} a \cdot \vec{s}
\end{equation}
where $g_{\text{aNN}}$ is the coupling constant, the pseudoscalar field $a$ is given by
\begin{equation}
    a = a_0 \cos{\left(\frac{m_{\text{DM}}c^2}{\hbar}t - \frac{\vec{p} \cdot \vec{x}}{\hbar}\right)},
\end{equation}
the magnitude of the field $a_0 = \sqrt{2\rho_{\text{DM}}}/(m_{\text{DM}} c)$, and the axion-like particle momentum is $\vec{p}$.  This interaction, which has the same form as a magnetic interaction, gives rise to an energy shift between states with opposite spin direction.  Assuming that the axion-like particle momentum is determined by its virial velocity, the energy shift is given by \cite{Graham2018}
\begin{equation}
   \Delta E = g_{\text{aNN}}\sqrt{2\rho_{\text{DM}}}\,\Delta m\, \frac{v}{c}\cos{\left(\frac{m_{\text{DM}} c^2}{\hbar} t\right)}
\end{equation}
where $\Delta m$ is the difference in the spin angular momentum projection and $v = 10^{-3}c$ is the virial velocity.

\begin{figure}
    \centering
    \includegraphics[width=0.40\textwidth]{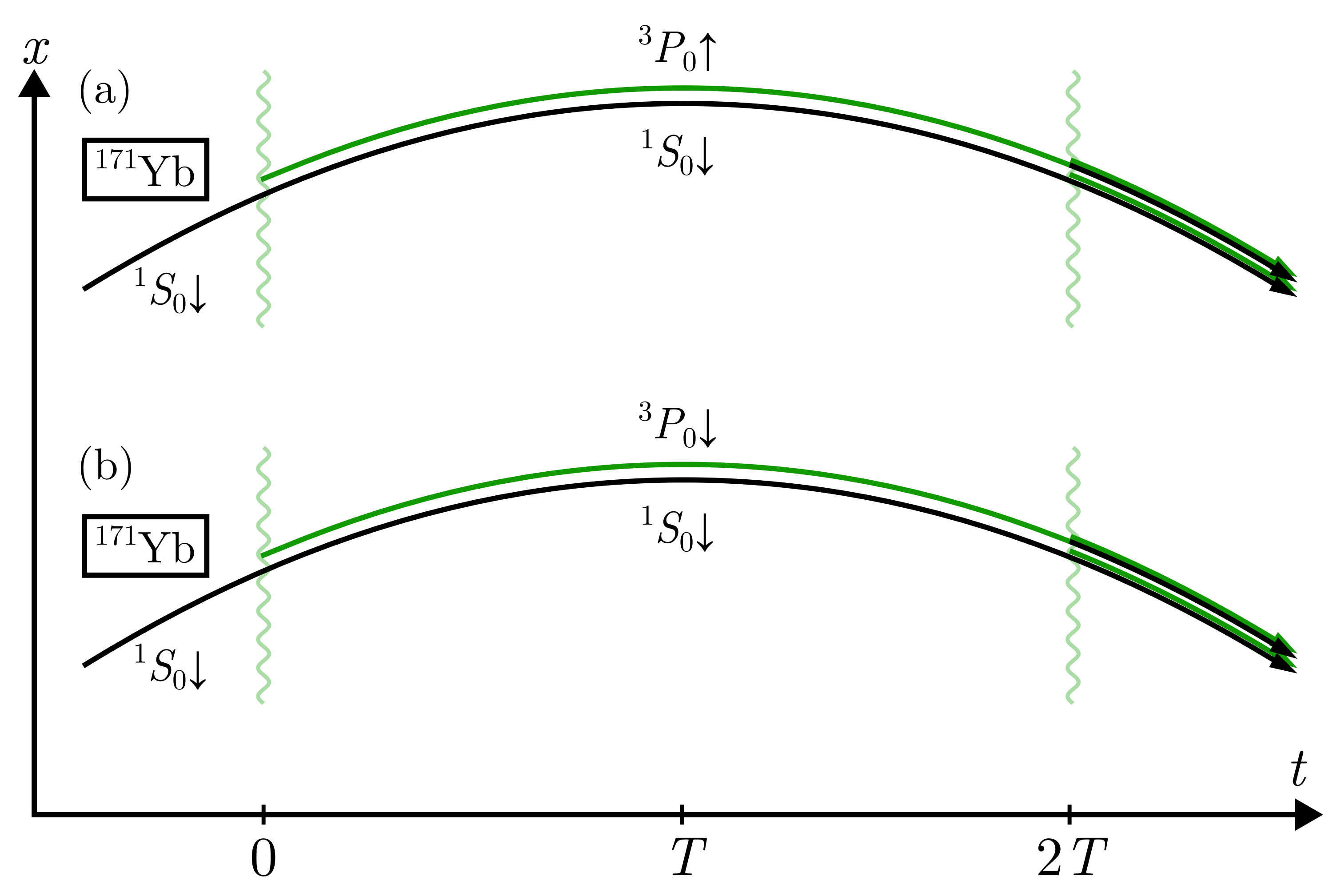}
    \caption{Interferometer geometry for searching for spin torque due to axion-like dark matter.  Doppler-free two-photon transitions at time $t = 0$ create superpositions of the $^1S_0$ electronic ground state and the $^3P_0$ excited state.  The differential phase between the $\Delta m = 1$ interferometer (a) and the $\Delta m = 0$ interferometer (b) is sensitive to spin torque but insensitive to the electronic transition frequency.  To reduce systematic effects, the two interferometers are spatially overlapped; interferometers are spaced vertically in the diagram for clarity.}
    
    \label{fig:Two-photon_spin-torque}
\end{figure}

\begin{figure}
    \centering
    \includegraphics[width=0.40\textwidth]{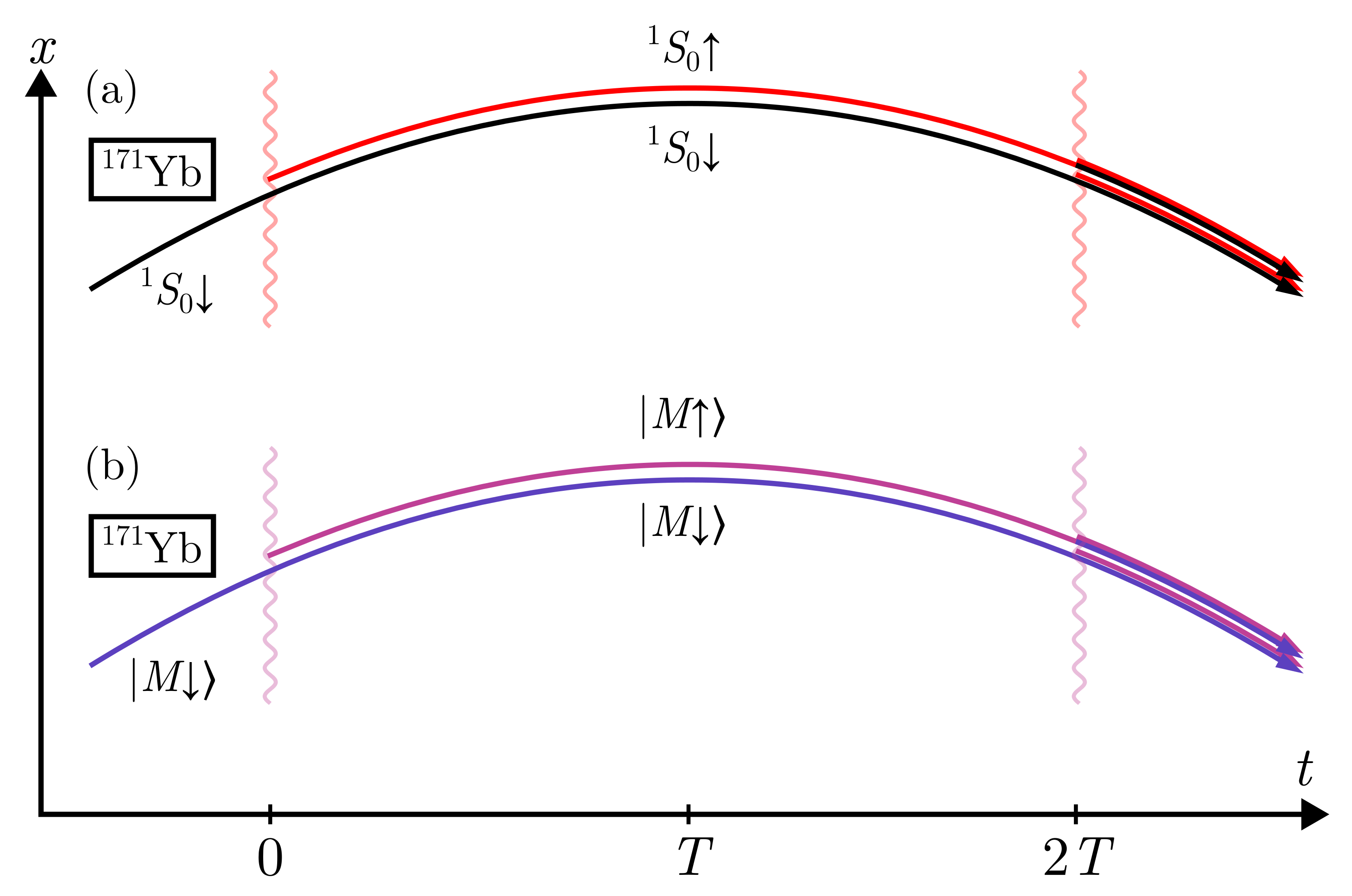}
    \caption{Alternative interferometer geometry for searching for spin torque due to axion-like dark matter.  In (a), the atoms are initialized in the electronic ground state, and a Ramsey interferometer is implemented between hyperfine ground state levels.  In (b), the Ramsey interferometer is instead carried out between magnetically sensitive states $\ket{M \uparrow}$ and $\ket{M \downarrow}$ (e.g. in the metastable $^3P_2$ level).  The interferometer in (a) is sensitive to spin torque but relatively insensitive to the magnetic field, while the interferometer in (b) measures the magnetic field to characterize its systematic effect.}
    
    \label{AC_field_spin-torque}
\end{figure}

Atom interferometers can be sensitive to this pseudoscalar dark matter coupling by searching for time dependence in the frequency of any transition that flips a nucleon spin.  Following the proposal of Ref.~\cite{Graham2018}, an interferometer geometry that could be used for such a measurement is shown in Fig.~\ref{fig:Two-photon_spin-torque}.  This configuration is sensitive to spin torque via the interferometer with the $\Delta m = 1$ transition, while the interferometer with the $\Delta m = 0$ transition is used to suppress systematic effects such as laser frequency drift.  An alternative approach is illustrated in Fig.~\ref{AC_field_spin-torque}.  Here the Ramsey interferometer between the hyperfine ground states of $^{171}$Yb is sensitive to the pseudoscalar field, while the second Ramsey interferometer between magnetically sensitive states (e.g. states in the metastable $^3P_2$ manifold) is used as a comagnetometer to decorrelate phase shifts from the magnetic field.  We note that $^{171}$Yb, which has nuclear angular momentum $I = 1/2$ from the spin of a single unpaired neutron, provides a simpler platform for this measurement than $^{87}$Sr \cite{Graham2018}, which has $I = 9/2$ from a combination of spin and orbital angular momentum.

\begin{figure}
    \centering
    \includegraphics[width=0.45\textwidth]{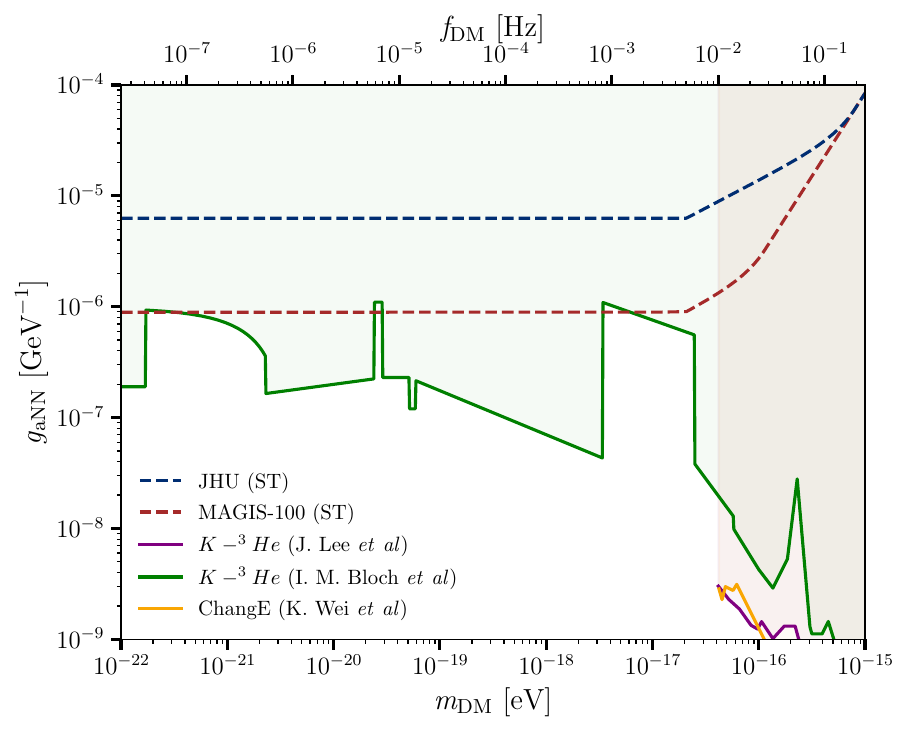}
    \caption{Constraint plot and projected sensitivities to axion-like dark matter coupled to nuclear spin.  JHU (ST):  projected sensitivity of spin torque search in JHU apparatus with interferometer geometry shown in Fig. 7 or Fig. 8.  MAGIS-100 (ST):  projected sensitivity of spin torque search with Yb in MAGIS-100.  Also shown are existing constraints from atomic magnetometry experiments (Bloch et al. \cite{Bloch2020}, green curve; Lee et al. \cite{Lee2023}, purple curve; Wei et al. \cite{Wei2024}, orange curve).  Shaded regions are excluded by existing constraints.  All curves are 95\% confidence limits.}
    
    \label{fig:constraints_spin_torque}
\end{figure}

The projected sensitivity of the JHU apparatus to axion-like dark matter through the axion-nucleon coupling is shown in Fig.~\ref{fig:constraints_spin_torque}.  We also plot existing bounds from atomic magnetometer experiments \cite{Bloch2020, Lee2023} and the projected sensitivity of a spin torque search utilizing $^{171}$Yb in MAGIS-100, assuming the same phase resolution as in the JHU apparatus and an interferometer time of $T = 4.5$~s.  Although the laboratory-scale interferometer is not expected to be competitive with the magnetometer limits, a long-baseline interferometric measurement would have comparable sensitivity in several regions of parameter space.  Our projections are more conservative than those reported in previous work \cite{Graham2018} due to our more conservative phase resolution estimates.

\section{IV.  Control of systematic effects}
\label{sec:Systematics}

Next, we consider the main systematic effects affecting these dark matter searches.  Since the measurements outlined in the previous section will be sensitive to many of the same systematic effects, control of a given systematic effect can benefit multiple dark matter searches simultaneously. 

The spectroscopic search for scalar dark matter [Figs.~\ref{fig:Alpha_time} and \ref{fig:Two-photon_alpha_time}; projected sensitivity curve ``JHU $\alpha(t)$'' in Fig.~\ref{fig:constraints_alpha_variation}] relies on a differential frequency measurement between the $^1S_0 \leftrightarrow {}^3P_0$ and the $^1S_0 \leftrightarrow 4\text{f}^{13}6\text{s}^{2}5\text{d}\,(J = 2)$ transitions in $^{171}$Yb.  Thus, the measurement is susceptible to differential frequency shifts in the lasers used to drive these transitions.  To suppress this noise source, the two lasers will be locked to the same optical cavity so that laser frequency drifts are common to the two transitions.  The optical cavity can utilize crystalline reflective coatings \cite{Cole2016} to reduce the influence of thermal noise in the coatings on the effective cavity length. 

The accelerometric searches for scalar and vector dark matter [Fig.~\ref{fig:Isotope_gradiometer}; projected sensitivity curves ``JHU (DC)'' and ``MAGIS-100 (DC)'' in Figs.~\ref{fig:constraints_alpha_variation} and \ref{fig:constraints_acceleration}, projected sensitivity curves ``JHU (AC)'' and ``MAGIS-100 (AC)'' in Fig.~\ref{fig:constraints_acceleration}] are sensitive to differential forces between the two Yb isotopes.  Based on previous atom-interferometric equivalence principle tests \cite{Asenbaum2020a}, the leading systematic effect of this kind is expected to arise from AC Stark shifts induced by the interferometry lasers.  AC Stark shifts can be managed by utilizing a compensated optical spectrum \cite{Kovachy2015b, Asenbaum2020a} and by controlling the intensity, size, and divergence of the beams.  We note that static equivalence principle tests are much more challenging from the perspective of systematic errors than searches for oscillating signals, as the static tests are subject to DC systematic shifts from magnetic fields, gravity gradients, and other sources \cite{Asenbaum2020a}. 

The largest source of systematic error for the pseudoscalar search [Figs.~\ref{fig:Two-photon_spin-torque} and \ref{AC_field_spin-torque}; projected sensitivity curves ``JHU (ST)'' and ``MAGIS-100 (ST)'' in Fig.~\ref{fig:constraints_spin_torque}] will likely be the magnetic field, since the frequency of a $\Delta m = 1$ transition is inherently sensitive to the magnetic field amplitude at first order.  The magnetic field in the interferometer region will be controlled with a multiple-layer mu-metal shield \cite{Wodey2020}, and the magnetic field along the interferometer trajectory will be measured {\it in situ} by using states with higher magnetic sensitivity \cite{Asenbaum2020a}.

Several systematic effects will affect all three dark matter searches.  For example, initial position and velocity displacements between interferometers produce phase shifts in the presence of gravity gradients.  These phase shifts can be suppressed by overlapping the interferometer midpoints and by adjusting the frequencies of the interferometer lasers during the sequence.  The idea of adjusting laser frequencies to compensate for vertical phase gradients, first proposed in Ref.~\cite{Roura2017}, has been experimentally demonstrated in dual-isotope precision measurements \cite{Overstreet2018,Asenbaum2020a}.  Crucially, this approach naturally improves as the sensitivity of the experiment improves, greatly relaxing constraints on how well the two isotopes must be overlapped spatially.  Nevertheless, the initial position and velocity of each interferometer will need to be controlled at the level of $50~\mu$m and $50~\mu$m/s, respectively \cite{Asenbaum2020a}.  The Coriolis effect, which causes a velocity-dependent phase shift, can be reduced by counter-rotating the retroreflection mirror so that the interferometer laser direction remains constant in the freely falling frame of the atoms \cite{Dickerson2013,Lan2012}.  We note that horizontal phase gradients from any source can be suppressed by appropriate changes to the angle of the retroreflection mirror during the interferometer sequence \cite{Asenbaum2020a}.  As with vertical phase gradients, this compensation naturally improves with the sensitivity of the experiment.  Shifts from blackbody radiation \cite{Haslinger2018} can be managed by controlling the temperature of the interferometry region.

\section{V.  Discussion and conclusion}
\label{sec:Discussion}

Ytterbium atom interferometry in atomic fountains can be used to perform sensitive searches for dark matter.  The projected sensitivity to variations of the fine-structure constant is especially notable, as a differential frequency measurement between ytterbium electronic transitions is expected to surpass previous constraints on $d_e$ from ytterbium ion clocks by two orders of magnitude in the dark matter mass range from $10^{-22}$ eV to $10^{-16}$ eV.  This improvement is made possible by combining the lower quantum projection noise of neutral-atom systems with the high sensitivity of open-f-shell electronic states to variations of the fine-structure constant.  Compared to a proposed ytterbium optical clock \cite{Safronova2018}, the JHU atom interferometer is expected to have similar sensitivity while avoiding systematic effects from lattice light shifts and the density shift.  

The most stringent search for vector dark matter with ytterbium atom interferometry can be derived from a static equivalence principle test between different atomic species, which is naturally sensitive to the static forces that would arise from a dark matter field sourced by the Earth.  Since the vector coupling charge-to-mass ratios are similar for ytterbium isotopes and for the test masses used in the MICROSCOPE experiment \cite{Touboul2022}, an improved search requires performing an atom-interferometric equivalence principle test with a relative accuracy better than $10^{-15}$.  Such a test is challenging but possible in laboratory-scale devices and is feasible with long-baseline experiments such as MAGIS-100.  A more accurate equivalence principle test would also provide improved sensitivity to the scalar dark matter couplings parameterized by $d_e$, $d_{m_{e}}$, $d_{\hat{m}}$, and $d_{\delta m}$ \cite{Damour2010}.  Searches for time-varying acceleration induced by vector dark matter can be implemented as well.  In MAGIS-100, this type of measurement could enable an improved search for vector dark matter without requiring the control of DC systematic effects. 

Atom interferometers can search for axion-like dark matter through the axion-fermion coupling, which induces spin torque.  The $^{171}$Yb isotope, which has a $^1{S}_0$ electronic ground state and a nuclear angular momentum $I = 1/2$ derived from a single unpaired neutron, is technically ideal for this experiment.  We find that a spin torque search with $^{171}$Yb in MAGIS-100 would be competitive with existing limits from atomic magnetometers in some mass ranges.  

Finally, we note that the experimental parameters used to generate sensitivity estimates in this work (atom number, beamsplitter momentum transfer, etc.) are based on values that have previously been achieved in atom interferometers.  Anticipated improvements in phase resolution via increased atom number, decreased cycle time, or the use of a squeezed atom source \cite{Hosten2016} will lead to corresponding increases in dark matter sensitivity.  Likewise, the coherence time can be further increased in long-baseline detectors \cite{Abe2021, Badurina2020, Canuel2018}, trapped atom interferometers \cite{Xu2019}, or space-based experiments \cite{El-Neaj2020}. 

\section{Acknowledgments}

We thank Wei Ji, Jia Liu, and Ken Van Tilburg for useful discussions.  M.P. acknowledges support from a William H. Miller III Graduate Fellowship.

\section{Appendix}

\subsection{A.I. Charges and couplings} \label{app:charges}
The charge of an atom under the scalar field-electromagnetic field coupling, as discussed in Sec. \hyperref[sec:Scalar]{III.A}, is given by \cite{Damour2010}
\begin{equation}\label{eq:dilaton charge}
    Q_e=\frac{Am_N}{m_A}\left(-1.4+8.2\frac{Z}{A}+7.7\frac{Z(Z-1)}{A^{4/3}}\right)\times 10^{-4}
\end{equation}
where $m_N$ is the mass of a nucleon and $m_A$ is the total mass of the atom. Similarly, when considering the scalar coupling to the electron mass, as mediated by the interaction term
\begin{equation}
    \mathcal{L} \supset d_{m_e}\, \varphi\, m_e c^2\, \bar\psi_e\psi_e,
\end{equation}
the resulting charge of an atom is expressed as \cite{Damour2010}
\begin{equation}
    Q_{m_e}= 5.5\times10^{-4}\, \frac{m_N}{m_A}\, Z.
\end{equation}
Finally, as discussed in Sec. \hyperref[sec:Scalar]{III.B}, the charge of an atom for a vector coupling to $B - L$ is given by the neutron number,
\begin{equation}
    Q_{B - L} = A - Z.
\end{equation}

Table I presents the relevant charge values for each coupling.

\begin{table}[h!]
    \centering
    \begin{tabular}{c c c}
        \hline
        \hline
        \multicolumn{1}{c}{Dimensionless} & \multirow{2}{*}{$Q\left(^{170}\text{Yb}\right)$} & \multirow{2}{*}{$Q\left(^{176}\text{Yb}\right)$} \\ 
        \multicolumn{1}{c}{Coupling} & & \\ 
        \hline
%         $g_{aNN}$ & - & - \\  
         $d_e$ & 0.004148 & 0.003958 \\  
         $d_{m_e}$ & 0.0002266 & 0.0002188 \\  
         $g_{B-L}\frac{1}{\sqrt{\hbar c}}$ & 100 & 106 \\  
        \hline
        \hline
    \end{tabular}
    \caption{Charges of ytterbium isotopes $^{170}$Yb and $^{176}$Yb for each of the dark matter couplings considered in this work.}
    \label{tab:charges}
\end{table}

\subsection{A.II.  Earth model} \label{app:earth}
The dark matter interactions that we consider in this work give rise to static forces between Standard Model particles of the form
\begin{equation}
    \vec{F} = \frac{d^2}{4\pi r^2}\,q_{1}\,q_{2}\, \left(1 + \frac{r}{\lambda} \right)e^{-r/\lambda}\, \hat{\mathbf{r}}
\end{equation}
where $\lambda = h/(mc)$ is the Compton wavelength of the dark matter field and $d$ is the relevant coupling constant. In the limit $m \to 0$, the force becomes Coulomb-like. It can be shown \cite{Adelberger2003, Hees2018} that the force on a test charge resulting from a uniformly-charged sphere of radius $R_S$ is
\begin{equation} \label{eq:F_Yukawa_sphere}
    \vec{F}_{\text{sphere}} = \frac{d^2}{4\pi r^2} q_{1} \tilde{q}_S\,\left(1 + \frac{r}{\lambda} \right)e^{-r/\lambda}\, \hat{\mathbf{r}}.
\end{equation}
Here $r$ is the distance from the center of the sphere to the test charge and
\begin{equation}
    \tilde{q}_S = q_{S} \Phi\left(\frac{R_S}{\lambda}\right),
\end{equation}
where $q_S$ is the total charge of the sphere and 
\begin{equation}
    \Phi(x)=3\frac{x \cosh{x}-\sinh{x}}{x^3}
\end{equation}
is a geometry-compensating function.

Because forces of the form $\vec{F}$ are linear functions of charge, complex distributions can be built by summing simpler ones. When finding the force on a test charge induced by the Earth, we make the substitution $\tilde{q}_S \to \tilde{q}_{\text{Earth}}$ in Eq.~\ref{eq:F_Yukawa_sphere} and model the Earth as an iron core surrounded by a thick silicon dioxide shell:
\begin{equation}
    \begin{split}
        \tilde{q}_{\text{Earth}} &= q_{\text{Fe}} \Phi\left(\frac{R_{\text{Core}}}{\lambda}\right)\\
        &+ q_{\text{Si}} \left[\Phi\left(\frac{R_{\text{Earth}}}{\lambda}\right) - \frac{V_\text{Core}}{V_\text{Earth}}\Phi\left(\frac{R_{\text{Core}}}{\lambda}\right)\right]\\
        &+ q_{\text{O}_2} \left[\Phi\left(\frac{R_{\text{Earth}}}{\lambda}\right) - \frac{V_\text{Core}}{V_\text{Earth}}\Phi\left(\frac{R_{\text{Core}}}{\lambda}\right)\right].
    \end{split}
\end{equation}
We set $R_{\text{Core}} = 1.22 \times 10^6$ m and $R_{\text{Earth}} = 6.378 \times 10^6$ m.  Furthermore, we model the Earth's mass as $33\%$ iron and $67\%$ silicon dioxide.

It is useful to express the extensive charges $q$ in terms of the intensive charges $Q$ appropriate to each species. In the case of the scalar coupling to the electromagnetic sector ($d_e$), the intensive charge is $Q_e$ (see Sec.\ \hyperref[app:charges]{A.I}), and the extensive charge is $q_i = \sqrt{4\pi G} m_i Q_e$, where $m_i$ is the total mass of element $i$. The same relationship holds for the scalar coupling to the electron mass $(d_{m_e})$ with intensive charge $Q_{m_e}$.  
For the case of the vector $B - L$ coupling $(g_{B-L})$, the intensive charge is $Q_{B-L} = A - Z$ and the extensive charge $q_i$ is simply $A - Z$ times the number of atoms of element $i$.  

\bibliographystyle{apsrev}
% \bibliography{library}

\begin{thebibliography}{72}
\bibitem{footnote1}
Note that any new ultralight boson would give rise to new static forces between Standard Model particles, regardless of whether the boson comprises the dark matter.
\bibitem{footnote3}
As in a gravimeter, the frequency of the laser will be chirped to keep the laser on resonance with the atomic transition for all four pulses.
\end{thebibliography}

\end{document}